\documentclass[aps,prl,twocolumn,superscriptaddress,showpacs]{revtex4}
\usepackage{epsfig}
\usepackage{changebar}
\usepackage{graphicx,array}
\usepackage{epsfig}
\usepackage{subfigure}
\usepackage{amsmath}
\usepackage{rotating}
\usepackage{amsfonts}
\usepackage{amssymb} 

\begin{document}

 
\title{Intermittent quakes and record dynamics in  the thermoremanent magnetization  of  a  spin-glass}  
\author{Paolo Sibani}
\email[]{paolo.sibani@fys.sdu.dk}
\affiliation{Institut for Fysik og Kemi, SDU, DK5230 Odense M, Denmark}
\author{G.~F.~Rodriguez}
\affiliation{Center for Magnetic Recording Research,
University of California, San Diego, CA, 92093}
\email[]{gfrodrig@gmail.com}
\author{G.~G.~Kenning}
\email[]{gregory.kenning@iup.edu}
\affiliation{Department of Physics, Indiana University of Pennsylvania\\
Indiana, Pennsylvania 15705-1098}
\date{\today}
\begin{abstract}
 A   method for analyzing  the intermittent 
behavior of  linear response data in aging systems
 is presented and applied to the 
spin-glass thermoremanent magnetization (TRM) data of Rodriguez et al. (Phys. Rev. Lett. 91, 037203, 2003).
 The   probability density function (PDF) of the magnetic fluctuations 
has  an asymmetric   exponential  tail, showing    that the 
 demagnetization process occurs through intermittent 
  spin rearrangements or \emph{quakes} which significantly  differ
from reversible fluctuations having  a Gaussian distribution 
with zero average. The intensity of quakes is determined by
the  TRM decay rate, which in turn depends on 
 $t$, the time  since the  initial  quench  and   on  $t_w$, the time  at which the 
magnetic field  is cut.  For a broad range of temperatures, 
these dependences  are  extracted numerically from the data and  
described  analytically using the assumption  that the system's 
linear response is fully  subordinated to the occurrence of the quakes  
which spasmodically release the imbalances  created by the initial quench. 
\end{abstract}


\maketitle
\section{Introduction} Many  glassy  materials  unable to 
re-equilibrate   after a  temperature quench 
undergo     \emph{aging},   a dynamical process  
whose key  properties are largely system  independent. 
In spin-glasses, a class of disordered magnetic materials,
one  thoroughly investigated aging quantity 
is  the  thermoremanent magnetization (TRM)~\cite{Alba86}.
In a TRM experiment,  a sample undergoes a rapid thermal  quench at $t=0$,
with a  small  magnetic field present. Subsequently,  the  temperature remains
constant throughout the experiment, while the magnetic field  is turned off
in a single step at  $t=t_w$. The decay of the  TRM  for  $t>t_w$  
 depends mainly on  the  scaling  variable  $(t-t_w)/t_w^\mu$,  where $\mu$ is  
a parameter close to unity~\cite{Alba86}.   
As  recently   recognized~\cite{Zotev03,Rodriguez03}, 
the value of $\mu$    depends  on  the  quenching  rate,
and approaches a so-called  full aging limit, $\mu = 1$,  as the 
rate   increases toward infinity, i.e. in  the limit of  an instantaneous
quench. These phenomena  together reveal  a persistent  memory of the  initial quench 
 and a strong sensitivity to the rate of cooling.  Memory 
 behavior  includes a number of   other fascinating  aspects~\cite{Jonason98,Sibani04a},
 and could be  rooted in a multi-scale, hierarchical  nature of the
energy landscape of  amorphous systems~\cite{Dotsenko85,Sibani89,Joh96,Hoffmann97,Bouchaud01}. 

Memory  issues can be further elucidated  by analyzing the
 fluctuations statistics  in meso-scaled
systems~\cite{Bissig03,Buisson03,Buisson04,Cipelletti05}.
 The basic observation is that the  Probability Density Function (PDF) of an aging quantity 
typically comprises  a  Gaussian part and an exponential tail.
The Gaussian covers reversible fluctuations with \emph{zero} average,
and the  tail  describes intermittent events, which carry the  drift of the aging process,
e.g., in the present case, they carry the  net change in magnetization.
The intermittent   events are  prominent  during the   non-equilibrium aging regime $t>>t_w$ where
the Fluctuation Dissipation theorem is clearly violated~\cite{Buisson03,Buisson04,Sibani05}.
Furthermore, as shown in the sequel,  they are equally prominent in a short time 
interval immediately following field removal. In general, the  statistical weight of 
intermittent fluctuations relative to the weight of Gaussian fluctuations
is determined by the rate $r_{TRM}$ of magnetization change, the latter depending on both
$t$ and   $t_w$ as detailed below.

Heat transfer statistics  directly  probes  thermally  activated 
dynamics, and is  easily collected in numerical 
simulation.  In  two different aging systems, the Edwards-Anderson 
 spin-glass model~\cite{Sibani05,Sibani04a} and 
a model with p-spin interactions~\cite{Sibani06b}, 
Gaussian and intermittent events  are  clearly identifiable. 
The idea that the intermittent events, or  \emph{quakes} 
are triggered by extremal, e.g.  record sized, reversible  
fluctuations~\cite{Sibani03,Sibani93a} explains the
age dependence of the PDF  of energy fluctuations.
The same `record dynamics'  approach 
was  applied to the configuration
 autocorrelation function~\cite{Sibani06} and to   dynamical properties of other complex 
systems~\cite{Anderson04,Oliveira05}. 

Arguably, quakes   influence 
aging   quantities other than  the energy, including e.g. 
the all important linear response functions. Unlike the energy,
response  functions   involve  the turning of  an external perturbation
on or off,  and hence require  additional  theoretical attention.
In \emph{equilibrium}   situations, it is well
known that a  small perturbing field  simply  probes   the spontaneous  fluctuations
of its conjugate variable. 
For out-of-equilibrium 
intermittency, an  idea similar in spirit can be formulated     as  a \emph{subordination}  hypothesis:
 significant configuration changes, whether
 field induced,  as in TRM,  or spontaneous, as in the de-correlation of
the   magnetization fluctuations of an unperturbed system,   
 only occur in conjunction  with the quakes.
These, in turn,   release the strain created by the initial quench  
in a temporal sequence which is, to linear order,  unaffected by the perturbation. 
This  subordination hypothesis was   recently   applied
to the autocorrelation function in the Edwards-Anderson spin-glass
in zero field~\cite{Sibani06}
and is  presently applied   to the experimental TRM decay.
At a price---short time effects of equilibrium like-fluctuations
are neglected---the approach  leads to a considerable mathematical simplification,  
as the statistical properties of e.g. the  response  flow  
from those of the quakes.

In the sequel,  we   first describe  the method of  data analysis.
We  then show the intermittency of the decay 
and   fit the age and time dependence of the TRM decay rate to
a simple formula.
The theoretical ideas behind this  formula  are spelled out in the Theory 
Section, and the whole paper is rounded off  with   a  brief 
concluding Section.

Last but not least, a   notational issue:   the
variable  `$t$'   denotes here   the time elapsed since the initial quench,
or `age',  and `$t_w$' 
denotes the specific  age at which  the external field is switched off. A
slightly different   convention, widespread 
 in the experimental literature,
see e.g. Refs.~\cite{Struik78,Vincent96,Rodriguez03},   uses
`$t$'  for the time elapsed after  field removal, a quantity 
presently denoted   by  `$t_{obs}$', with $t_{obs}=t-t_w$.
Furthermore, in refs.\cite{Sibani05,Sibani06a}, $t_w$ is used to 
denote the system age, our present $t$.  Our   choice 
emphasizes   the initial quench as the common origo 
for  all  time variables,   and  slightly simplifies the notation,
e.g. our scaling functions have a  $t/t_w$ rather than a  $1 + t/t_w$ 
argument.

\section{Method of data analysis}   In this Section,  the TRM data
 of  ref.~\cite{Rodriguez03} are analyzed  with focus on intermittency. 
In the experiment,   a  
   Cu$_{0.94}$Mn$_{0.06}$ sample, with critical temperature
$T_g= 31.5$K, is  rapidly cooled, with an effective cooling time
of $19$s. The sample is then aged isothermally at temperature $T$ 
in a weak magnetic field ($H=20 G$),
with the field removed at $t=t_w$. 
The TRM signal $M(t)$ is measured  for $t > t_w$  
and for different  values of $t_w$ and $T$:
$t_w = 50,100,300,630,3600, 6310$ and $10000$s and 
 $T=0.4, 0.6, 0.83, 0.9$ and $0.95$ $T_g$.
The    observation time $t_{obs} = t-t_w$ ranges  from 
appr. $6000$s to appr. $30000$s and 
the ratio $t/t_w$  correspondingly varies    from $1$  to  
$100$.  Measurements are  recorded  every   $1.045$s
for  $t_w <6310$s and every $2.04$s for
  $t_w \geq 6310$s.   
All  data are given as  dimensionless ratios of the TRM to the field 
cooled magnetization $M_{FC}$.

We consider  TRM   changes  $\delta M$  over a small time interval $\delta t$,
i.e. from the data we calculate   a time series  of magnetization differences 
  $\delta M (i)=   M(t_i+\delta t)- M(t_i)$,  
with   $t_i =  t_{i-1}+ \delta t$, and with $\delta t << t $ chosen
as a small multiplum    of the   measurement repeat time. 
  The  analysis has a  twofold aim: estimating  the PDF of the magnetization fluctuations 
 in different situations in order to show the presence of intermittency, and
estimating the time and age dependence of the  rate $r_{TRM}$ (henceforth 
simply 'rate') of magnetization decay, in order to compare with   theory. 
 
 The   PDF  are  straightforwardly  estimated by binning 
  the  $\delta M (i)$ values  
 sampled over suitable time intervals.  
 An estimate of $r_{TRM}$   is not easily obtained: 
 in order to uncover  the drift part
 of the dynamics, the   Gaussian  fluctuations, 
   which are far more frequent than intermittent events over small time intervals, 
 must be averaged out. Mainly, the averaging is  done   over  suitably
 chosen  subintervals of the
 observation interval.  Additionally, 
we   perform for $T= 0.83 T_g$  an   average over  an ensemble 
consisting of   data with the same $t/t_w$ value.    From this  ensemble we  
also construct the PDF of the magnetic fluctuations  over a short time interval 
near  $t_w$. 

For   time averaging,  we consider a set of   intervals 
$I_t=[(8/9)t,(10/9)t]$ with midpoints $t$   equidistantly 
spaced on a logarithmic scale. The interval   widths,   $0.22 t$,   are  chosen as
a     compromise between  the conflicting  requirements of    small  statistical error 
and    good temporal resolution.
Using the  values  available in each interval $I_t$,  the  average 
  $\mu_{\delta M}(t,t_w,\delta t)$
  and the variance $\sigma^2_{\delta M}(t,t_w,\delta  t)$ 
  of the magnetization change $\delta M$  are estimated at time $t$
using  standard formulas.
By varying  $\delta t$,  we ascertain that $\mu_{\delta M}$ is proportional 
to $\delta t$. The  proportionality constant, which 
 is identical to  $r_{TRM}(t,t_w)$, is estimated by linear regression.
The $t$ and $t_w$ dependence of $r_{TRM}$ thus obtained 
 is  fitted to  the  expression  predicted by record dynamics. As  a  final 
 step,   the dependence is  integrated with respect to $t$, leading to  an analytical 
 expression for the TRM decay,  which is compared with the original 
 data. The outcome of this  whole procedure is displayed in the 
 first  five panels of Fig.~\ref{big}. 
 
For ensemble averaging, we note that the $t/t_w$ dependence of 
the TRM  (full aging), which   is   established in the 
literature~\cite{Rodriguez03}  and confirmed by the present
analysis, implies  that   $\delta M$  values with the same 
$t/t_w$, i.e. $t/t_w = C$,  are physically equivalent.
Hence, they  can  meaningfully 
be collected   from   data streams
 taken at  different $t_w$ into ensembles
 labeled by the  value of $C$. 
This procedure is carried out  for the $T=0.83 T_g$ data,
collecting, in each case,  $\delta M$ values 
 within the interval $t \in [C t_w,1.25 C t_w]$ 
and systematically varying the value of $C$. 
From the $C = 1.13$ ensemble, we construct the  (unnormalized) PDF  of  $\delta M$ data.
We also estimate, for a number of different $C$, the rate 
 $r_{TRM}(t,t_w)$ as the  average of $\delta M/\delta t$  of the corresponding ensemble.
The results are  shown in  the main panel and the insert 
of Fig.~\ref{parameters}, respectively. 
\begin{figure*}[t]
\centering
\mbox{
\subfigure[ Intermittent and Gaussian PDF]{\includegraphics[width=.44\textwidth,height=.35\textwidth]{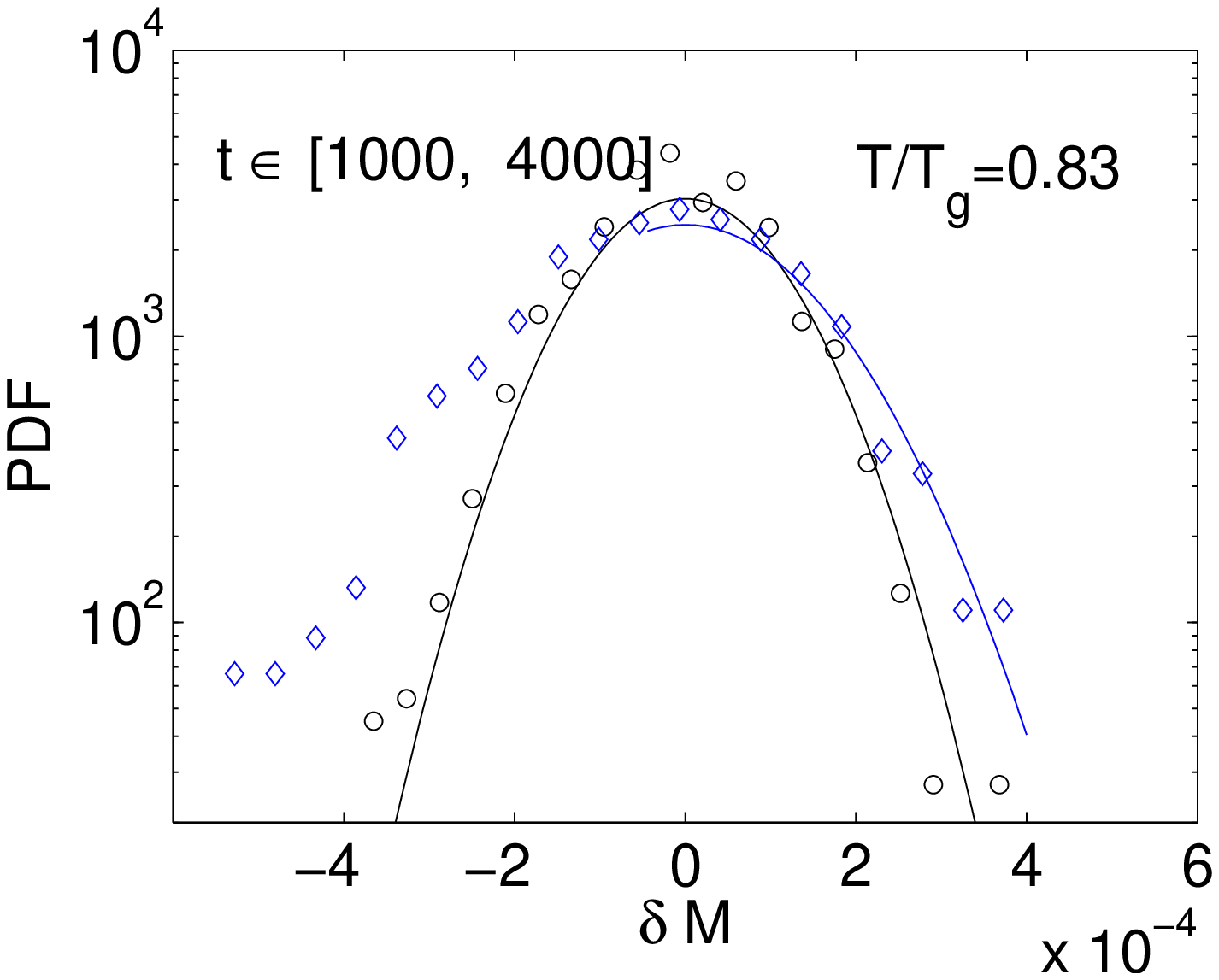}} \quad 
\subfigure[ Average and variance]{\includegraphics[width=.44\textwidth,height=.35\textwidth]{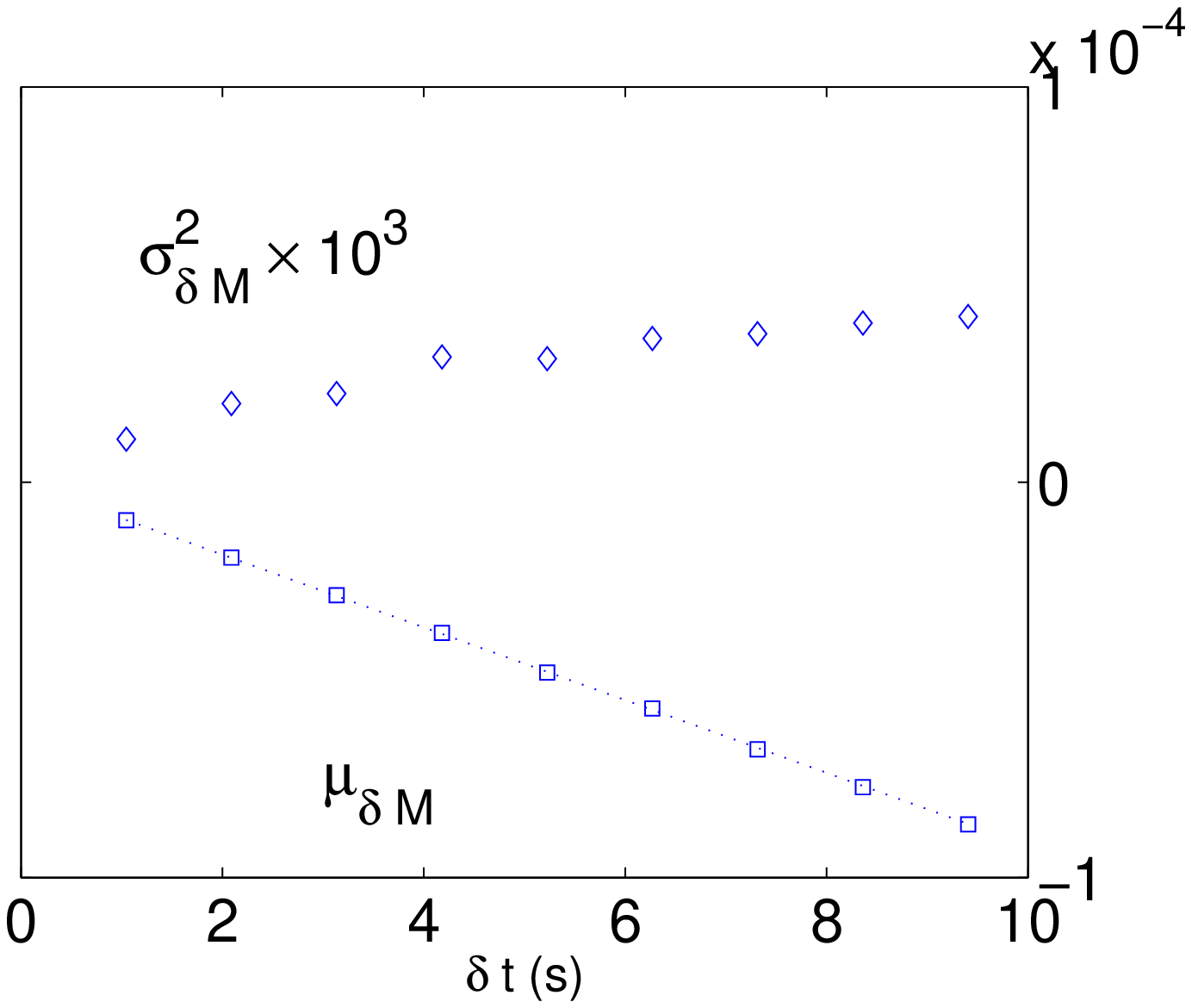}}} \quad 
  \quad  
\caption{ (Color on line)  {\em (a)}: The circles show the  PDF of the magnetization change $\delta M$
for isothermal aging at  $T=0.83 T_g$, with the field cut  at $t_w=100$s.
 The values of $\delta M$ are  taken over short intervals    $\delta t=1.045$s
within  the observation  interval    $[1000, 4000]$. Diamonds show 
the PDF of $\delta M$  from the same data stream, 
but now taken over larger intervals    $\delta t=3 \times 1.045$s.
This leads to a much stronger intermittent component on the left wing of the PDF. 
The full lines  are  least square fits   of the  \emph{positive} values of $\delta M$ to a Gaussian  with zero average. 
  {\em (b)}: From the same data, the   average and  the variance 
  (the latter scaled as shown) of  $\delta M$'s  are plotted versus $\delta t$.
The dotted line is obtained by a  least square error  fit. 
 } 
\label{PDFs}
\end{figure*}  
\section{Results} 
 Figure~\ref{PDFs}   illustrates
how  TRM  fluctuations  $\delta M$ occurring 
 over a small interval $\delta t$ have both a Gaussian component 
and an   intermittent  tail. The tail 
 becomes  more  dominant   as the \emph{average} change $\mu_{\delta M}$  
grows numerically larger. This differs from   
 `normal'  transport, where  increasing $\mu_{\delta M}$   would  only shift the center of the 
 Gaussian distribution  away from zero.  
 
Panel \emph{(a)} of Fig.~\ref{PDFs} compares two (unnormalized) PDFs of 
$\delta M$, which are both  obtained from  data    
with $T = 0.83 T_g$ and  $t_w=100$s. 
The  $\delta M(i)$  values   are collected over 
the same observation  interval $I=[1000s,4000s]$. 
 The PDFs are  shown on a logarithmic scale, 
where  a  Gaussian has a parabolic shape.
The nearly Gaussian PDF (circles) is for  $\delta t =1.045$s  and
the other PDF  (diamonds) is for $\delta t = 5\times 1.045$s. Increasing 
$\delta t$ increases the average magnetization change $\mu_{\delta M} = \delta t \, \,  r_{TRM}$ and 
hence increases the amount of intermittency.
Using   positive $\delta M$  values,
the central part of the PDF is fitted  (full lines) to a Gaussian
 with zero average.  Note how  the  Gaussian shape of the reversible
 fluctuations  remains visible  for positive $\delta M$ values,   in spite of the strong intermittent 
left wing present  for the larger $\delta t$ (diamonds).

 Panel \emph{(b)} of Fig.~\ref{PDFs}  shows 
the average  $\mu_{\delta M}$ (squares) and the variance $\sigma^2_{\delta M}$ 
(diamonds), plotted versus  $\delta t$.
As  $\mu_{\delta M}$ and $\delta t$ are expectedly  proportional, 
  the rate  of magnetization decay, (averaged over the observation interval $I$)
   can be  estimated as the
slope $r_{TRM}=\mu_{\delta M}/\delta t$.  
The dotted line   through the data
illustrates the quality of the linear fit. For completeness, we 
also show the variance $\sigma^2_{\delta M}$. As one would expect in a large
sample, $\sigma_{\delta M}$ is much smaller than the average. 

\begin{figure*} 
$ 
\begin{array}{cc}
\includegraphics[width=0.48\linewidth,height= 0.4\linewidth]{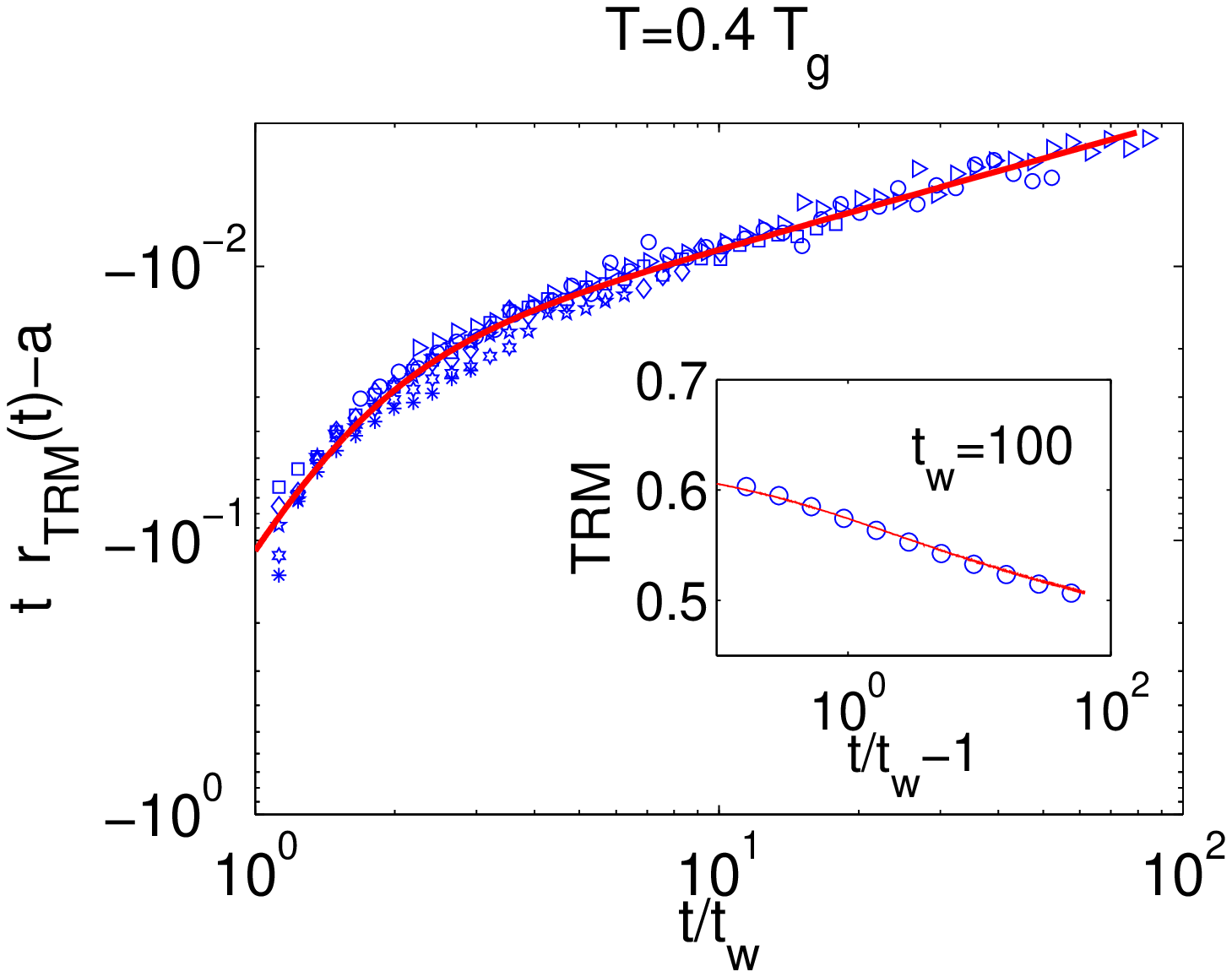}  &
\includegraphics[width=0.48\linewidth,height= 0.4\linewidth]{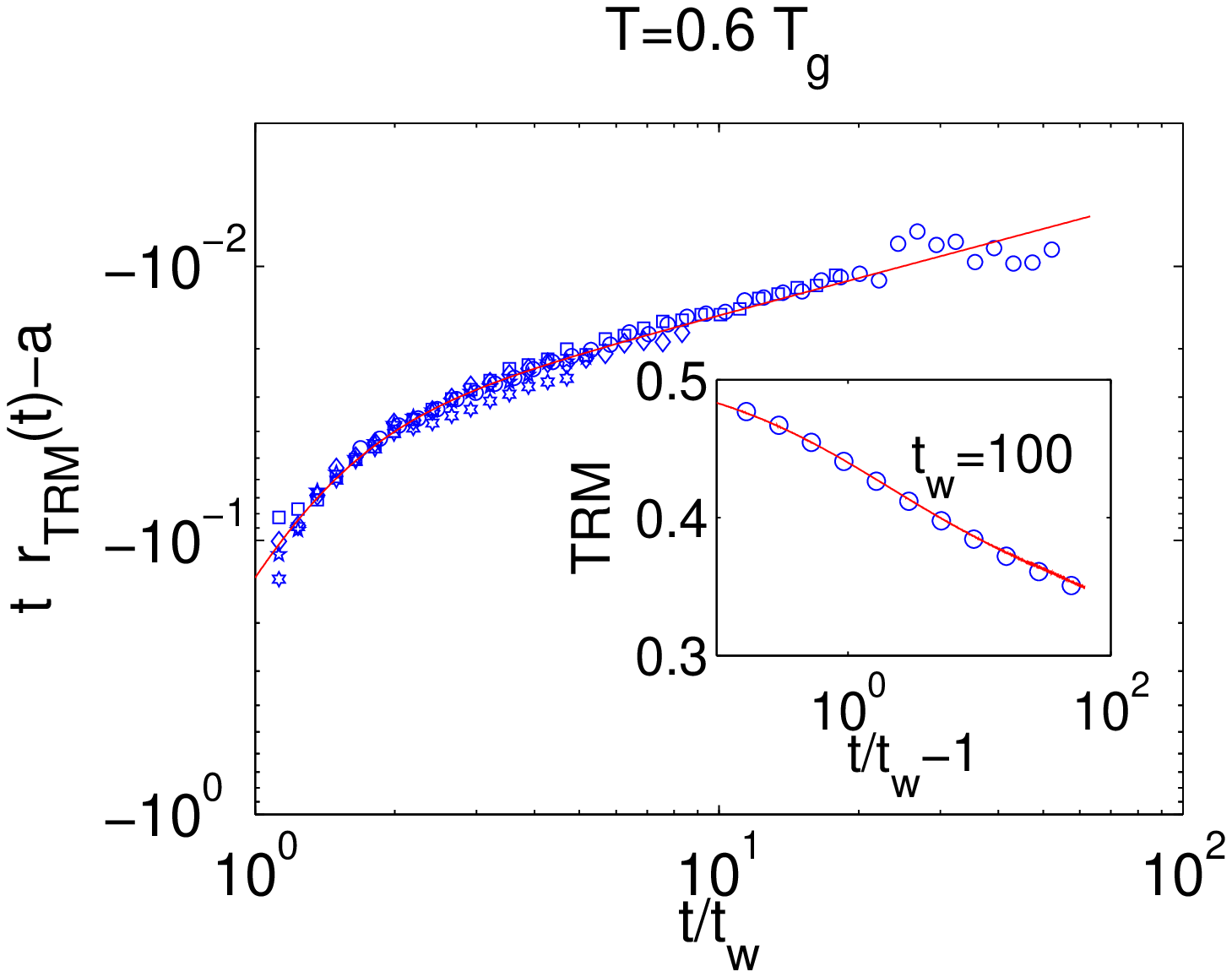} \\
 \includegraphics[width=0.48\linewidth,height= 0.4\linewidth]{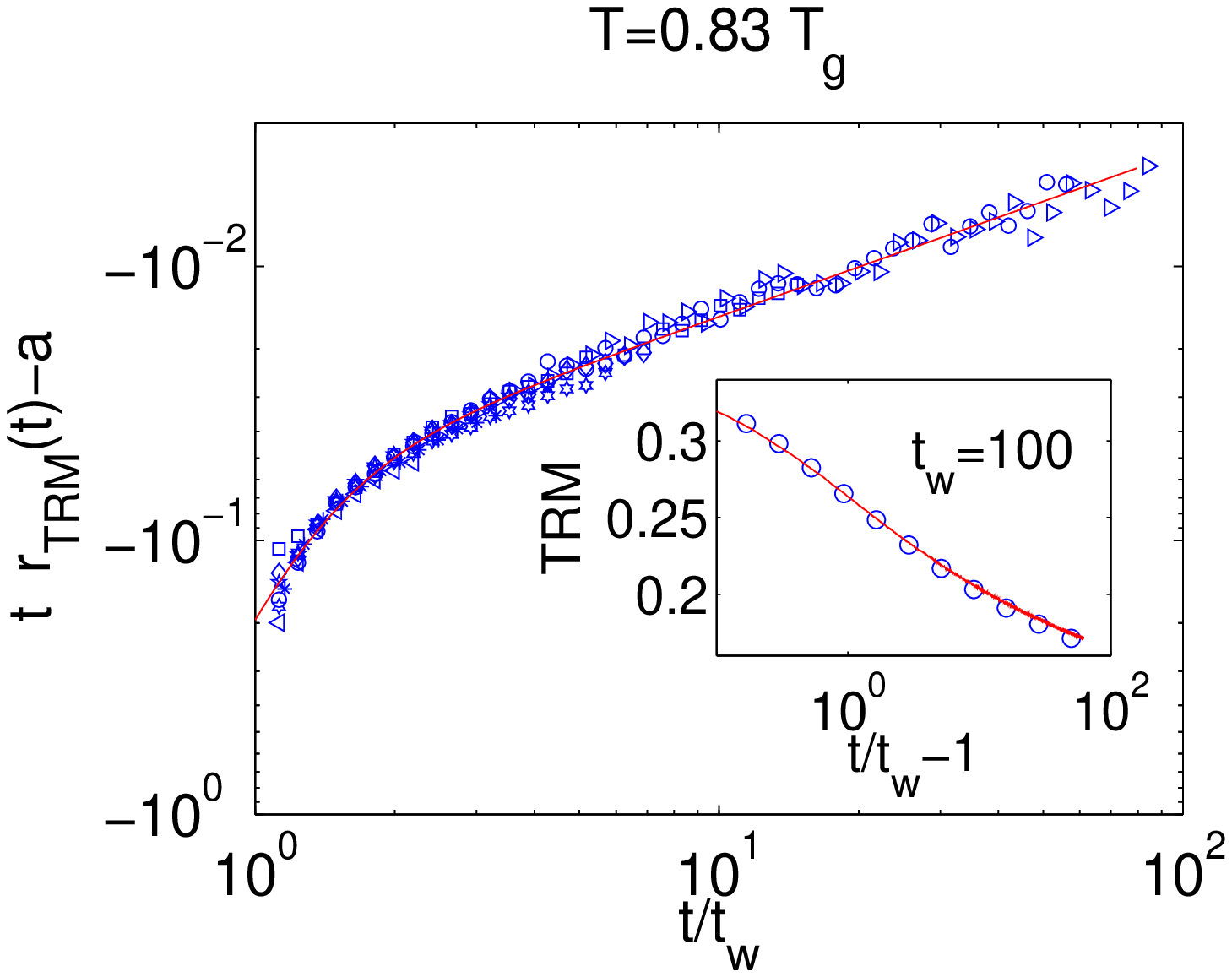}  &
 \includegraphics[width=0.48\linewidth,height= 0.4\linewidth]{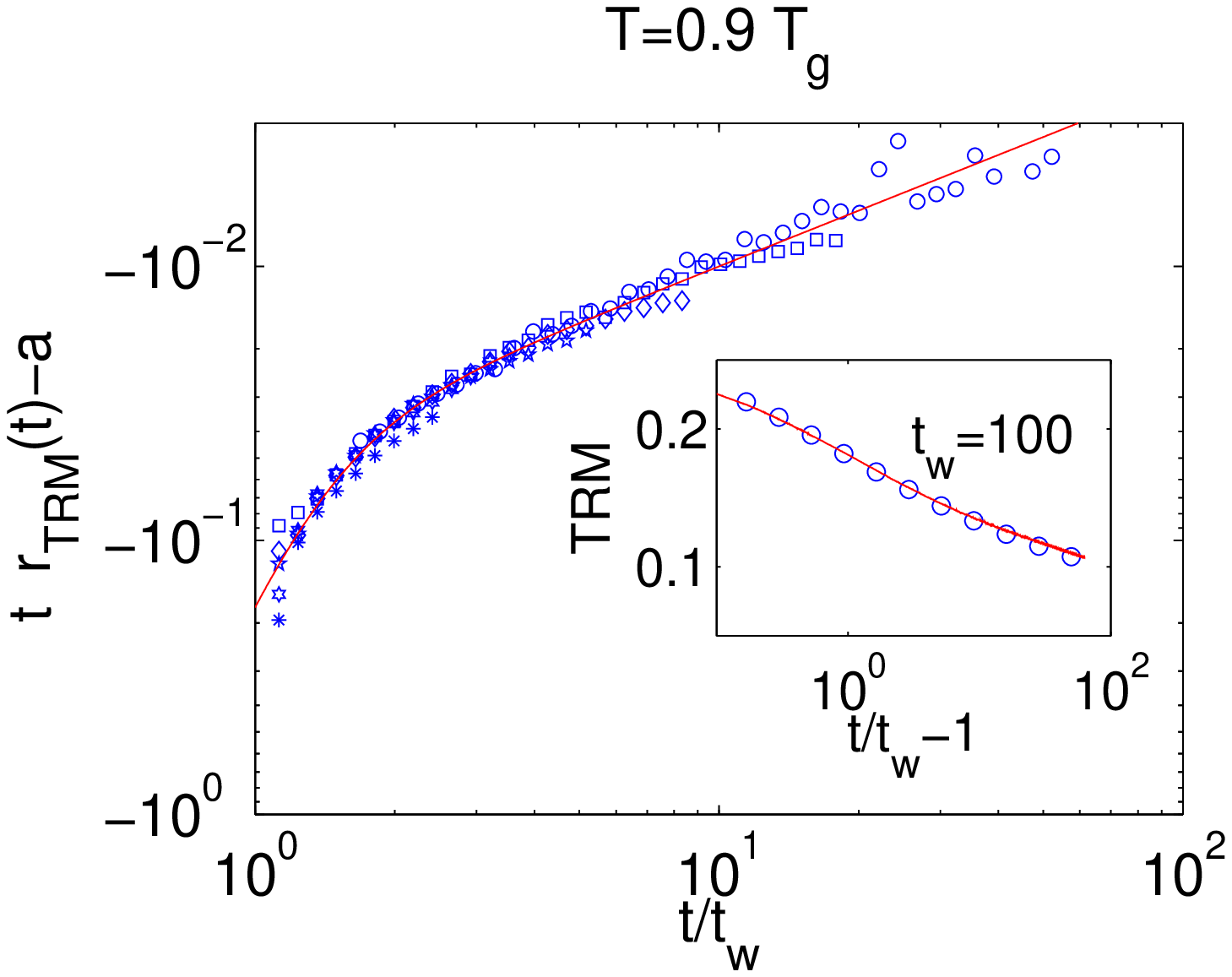}  \\
  \includegraphics[width=0.48\linewidth,height=0.4\linewidth]{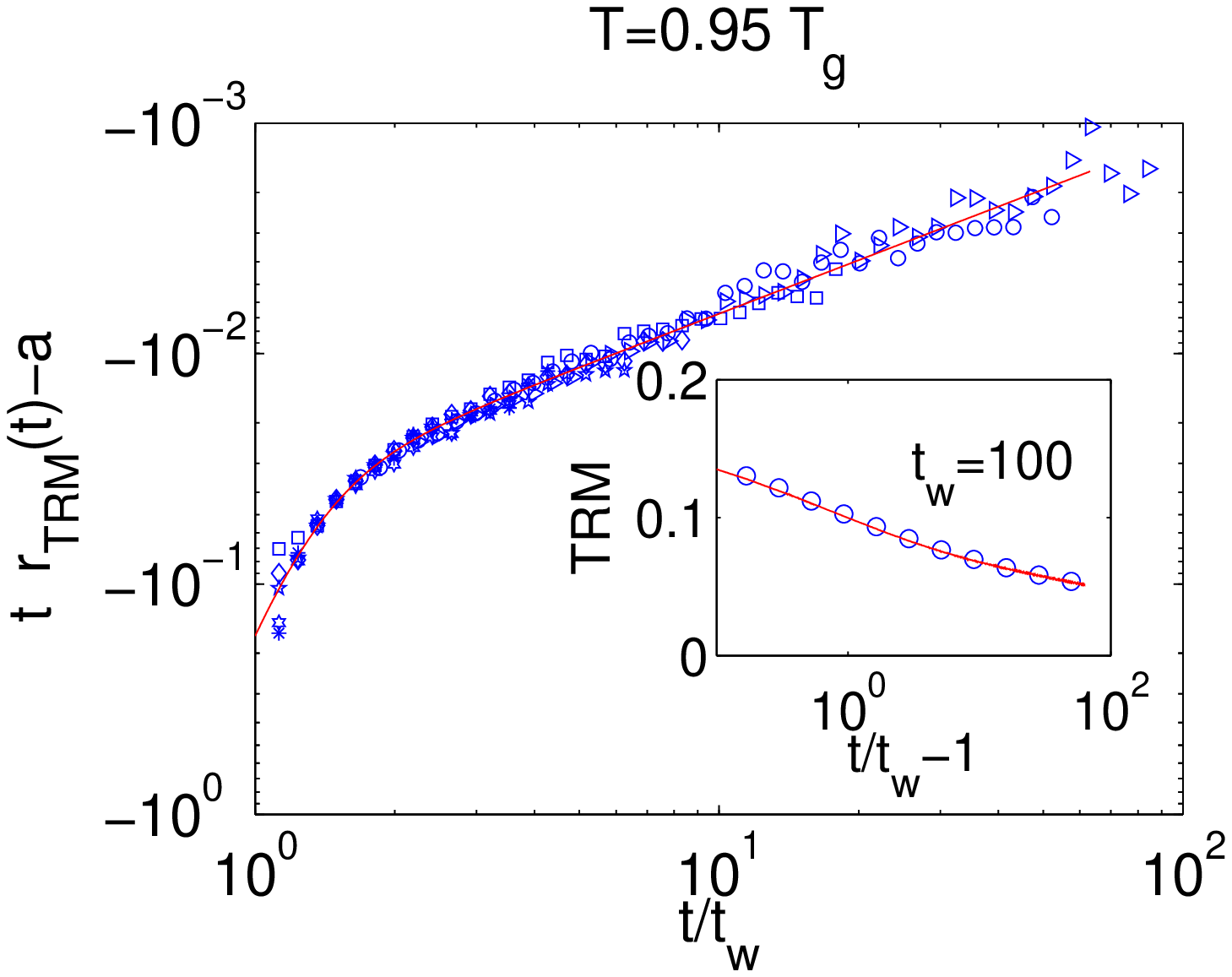}  &
 \hspace{0.8cm} \includegraphics[width=0.44\linewidth,height= 0.39\linewidth]{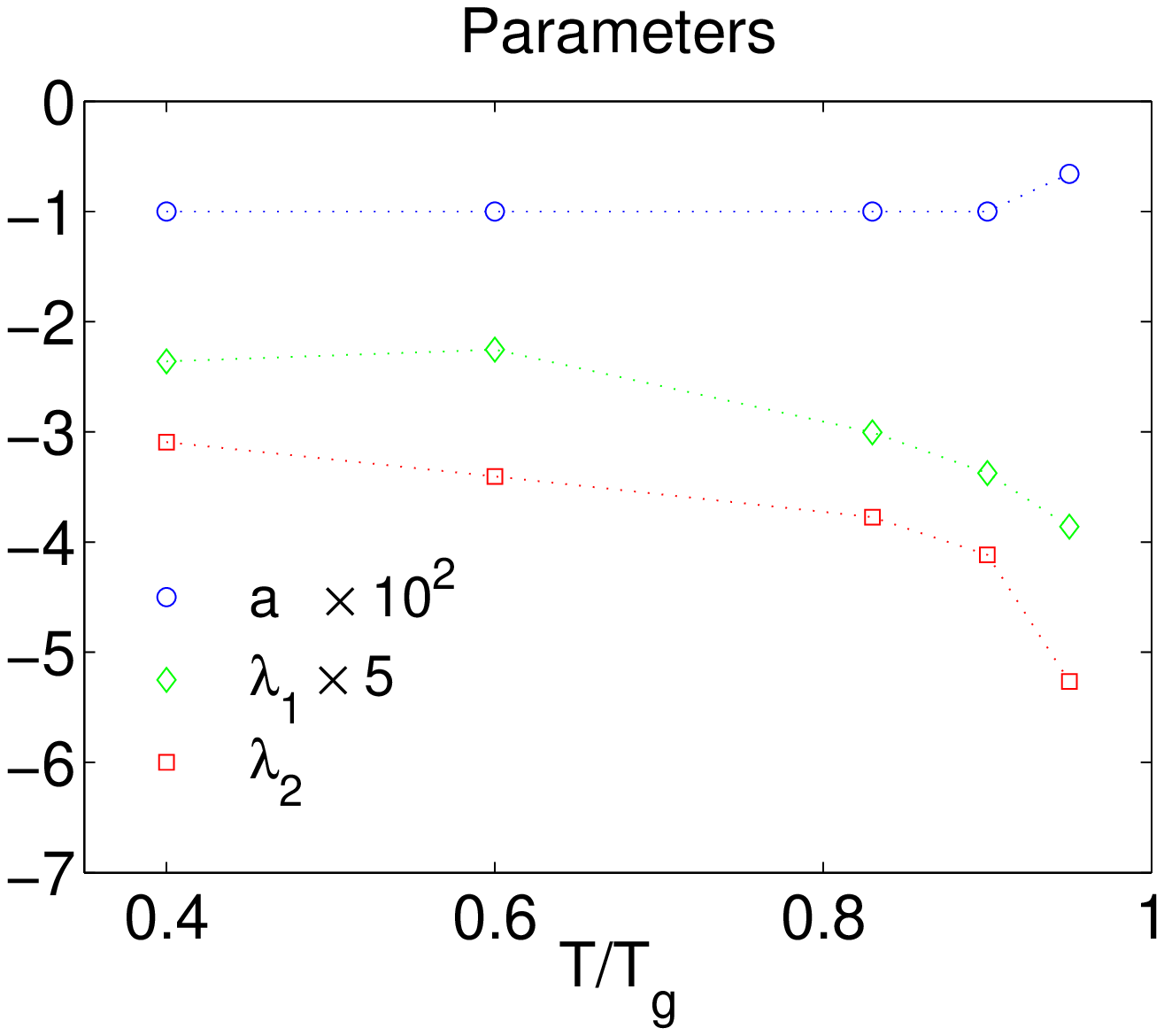}
\end{array} 
$  
\caption{ (Color on line) Panels 1-5: the main plots   show  the   estimated values 
of $ t \, \, r_{TRM}   -a $  versus $t/t_w$, for
$t_w= 50$ (right pointing triangles, unavailable for $T=0.6T_g$ and $T=0.9T_g$), 
$100$ (circles), $300$ (squares), $630$ (diamonds), $1000$ (pentagrams), 
$3600$ (hexagrams), $6310$ (asterisks)  and $(10000)$ (left pointing triangles).  
   The  lines are given by    Eq.~\ref{main_fit_formula}.
The inserts  compare  the    TRM decay measured at  $t_w=100$s (red line),  
with  the theoretical estimates (blue circles) obtained  by integrating the  
fitted   decay rate, see Eq.~\ref{finite_change}. The abscissa  $t/t_w-1$ is the ratio of the observation 
time to the field removal age. In   panel 6,  the parameter 
$a$ and the two exponents $\lambda_1$ and $\lambda_2$
are plotted vs. temperature. The lines are guides  to the eye.
}
\label{big}
\end{figure*}

 Panels 1-5  of Fig~\ref{big} illustrate 
 the  $t$ and $t_w$ dependence of the rate of magnetization
 $r_{TRM}$ for the five temperatures indicated. 
 Error bars  are obtained by standard methods and 
 data points with ($1\sigma$) uncertainty 
larger  than  $10$\%   are discarded. As it turns out, 
the rate approaches the functional form  $r_{TRM} \propto a/t$ for large values of $t/t_w$,
i.e. already  for $t/t_w > 10$ this is the  main contribution to $r_{TRM}$. 
Thus,   $t \, \, r_{TRM}-a$  plotted versus $t/t_w$  describes 
the   deviation of the rate from its  asymptotic behavior.  
The  full line  depicts the   function 
\begin{equation}
 y(t,t_w)  =     b_1 \left(\frac{t}{t_w}\right)^{\lambda_1} 
+b_2 \left(\frac{t}{t_w}\right)^{\lambda_2}.
\label{main_fit_formula}
\end{equation}
With   parameters  optimized  by a least 
square error fit, this function offers  a good analytical 
description of $t \, \, r_{TRM}  - a$. 
The  form of the parameterization  
is justified theoretically  in the next Section.
Here we note that  as  $\lambda_2 < -3$, the second term 
of Eq.~\ref{main_fit_formula}  only contributes  for  $t\approx t_w$, and 
that the sole  parameter of importance  in the limit $t >>t_w$  remains  $a$, which,
importantly,  remains  nearly 
 constant through the temperature range.

Let us  finally consider  the change in magnetization $\Delta M$ over an observation 
interval $I=[t_{i},t]$ starting 
at a time $t_i$  larger or equal to $t_w$, but otherwise arbitrary.
Plainly, 
\begin{eqnarray}
\Delta M(t_i,t,t_w) &=& \int_{t_i}^t r_{TRM}(t',t_w) dt' = a \log(t/t_i) \nonumber \\ 
&+&  \frac{b_1}{\lambda_1}\left[ \left(\frac{t}{t_w}\right)^{\lambda_1}-
\left(\frac{t_i}{t_w}\right)^{\lambda_1} \right] \nonumber \\  
&+& \frac{b_2}{\lambda_2} \left[ \left(\frac{t}{t_w}\right)^{\lambda_2}-
\left(\frac{t_i}{t_w}\right)^{\lambda_2} \right].
\label{finite_change}
\end{eqnarray}
For a generic observation interval, $\Delta M$ is a function of three variables,
as indicated. Customarily, one chooses $t_i=t_w$, and indeed,  
for  $t_i=t_w=100$s, the above formula yields  the analytical approximation $\Delta M(t,t_w)$
to the  TRM decay  plotted (blue circles) for selected values of $t$,
together with the measured  data (red line) in  the inserts of Fig.~\ref{big}. 
Conforming to standard usage,  the abscissa  $t/t_w-1$ is the ratio of the observation 
time to the field removal age.  Note that the $t_w$ dependent value of the magnetization at $t=t_w$,
formally an integration constant, is not provided by Eq.~\ref{finite_change}.
This  value   is determined     by   shifting $\Delta M(t,t_w)$   vertically,  
until  they best  overlap with the data is obtained. 
Finally,   the asymptotic decay of the magnetization  
only  acquires a  $t_w$ dependence if  $t_i=t_w$ is chosen. 
 
The last panel of Fig.~\ref{big} shows the temperature dependence
of the parameters $a$, $\lambda_1$ and $\lambda_2$.
Remarkably for thermally activated dynamics,  the  value $a=-0.01$ works for  all 
 temperatures except  for $T=0.95T_g$, i.e. very close to  the critical temperature.
 This behavior has a  simple interpretation:
 for $t/t_w$ sufficiently large, each quake contributes, on average, 
 the same amount to the TRM decay, and $a$ is proportional
to the logarithmic  rate of quakes, $\alpha$, which is $T$ independent
according to the theory.
  Finally,  the  temperature dependence of the two 
exponents $\lambda_1$ and $\lambda_2$ and of the 
corresponding pre-factors $b_1$ and $b_2$ 
(not shown),   is smooth and rather weak,  
  except near $T_g$.

\begin{figure}  
\vspace{-.5cm}
\center{
 \includegraphics[width=.48\textwidth]{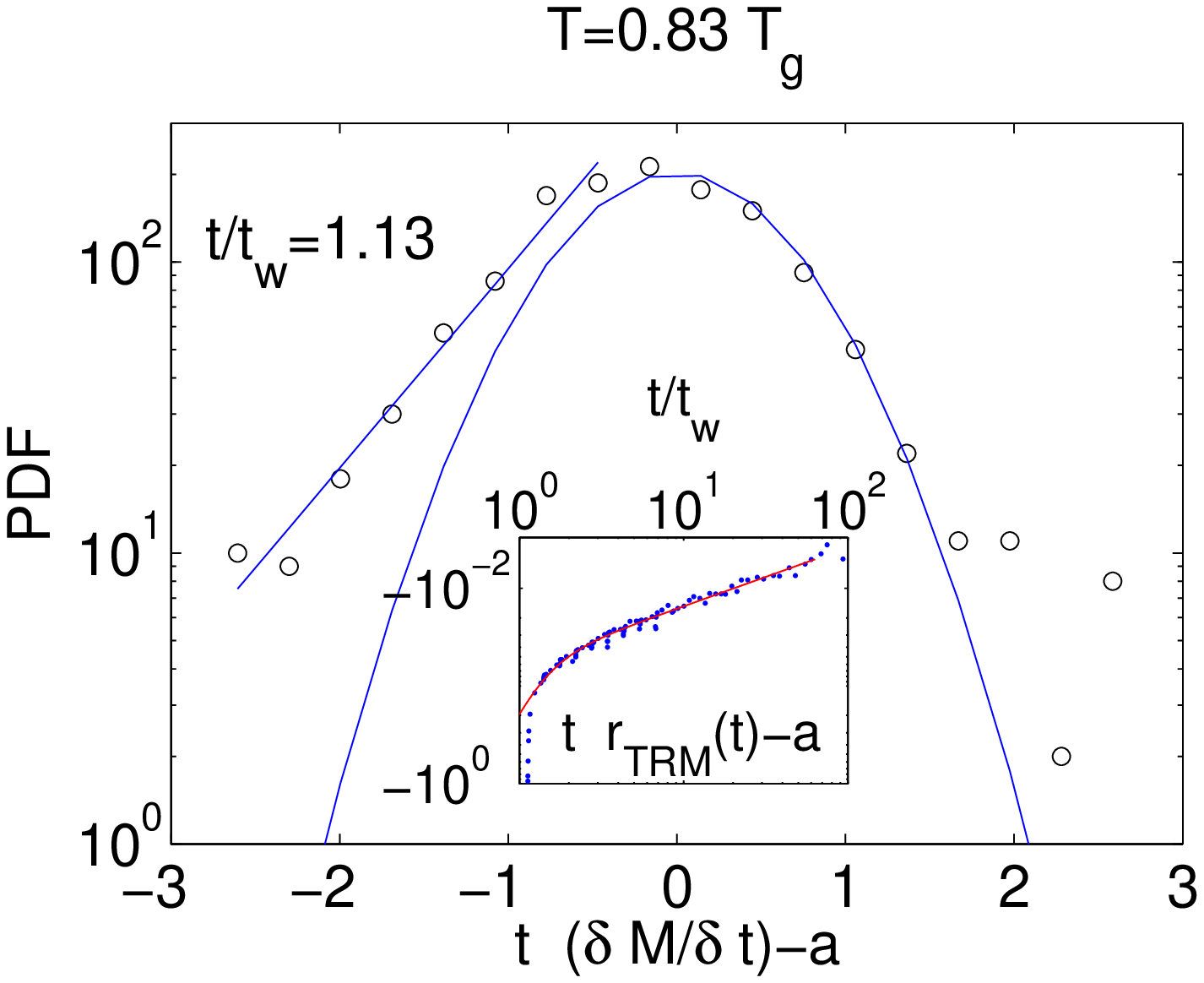}
 } 
\caption{
(Color on line) The main  plot   shows  the unnormalized  PDF (circles) 
 of the quantity  $\delta M/\delta t -a$. The values of  $\delta M$  
are  collected    from  data streams taken with  different
 $t_w$ and  with  $T=0.83 T_g$. The denominator   $\delta t$ is the repeat time in each 
measurement, All data are taken within  intervals  $[t, 1.25t]$, where the value of  $t$ 
changes with  $t_w$, keeping a constant ratio $t/t_w=1.13$.   The  
parabola and the straight line 
are  fits to a Gaussian and an  exponential, respectively. The  insert 
 shows  the  $t/t_w$ dependence of  $t \, \, r_{TRM} t -a$ (dots), where  
 $r_{TRM}$  is  estimated  by 
 averaging  $\delta M/\delta t$ over the   ensemble of data points available at each
 $t/t_w$.   
  The full line, which   is  lifted from   the third
 plot of Fig.~\ref{big},  is  obtained using  a different  averaging procedure,
 and  is therefore   \emph{not} a direct fit to the data shown.} 
\label{parameters}
\end{figure}

A property not  previously noticed  is
the  strong  intermittency    for $t \approx t_w$.
This  is  seen in  the main plot of Fig.~\ref{parameters}, where 
the   (unnormalized) PDF (dots) of $t \delta M/\delta t-a$ is  collected
 from intervals $[t,1.25t]$ with  $t/t_w = 1.13$, using all  available data streams
 with $T=0.83T_g$.   The denominator  $\delta t$ is the repeat
 time of the measurement. 
The PDF  features   a Gaussian  component and a strong intermittent component. as highlighted
 by  fits (full lines) to a  Gaussian with zero average (obtained from the  positive data
values),  and  an exponential  (obtained from the  negative data values).  
As a consistency check, we  estimate the quantity  
$t \, \, r_{TRM} -a$  as a function of $t/t_w$. The insert 
shows this quantity (dots) with $r_{TRM}$ estimated for each $t/t_w$ as the average of
 $\delta M/\delta t$ over the corresponding ensemble. The line---taken
 from the   third panel of Fig.~\ref{big}---is a fit 
 obtained  via time averages.
The    agreement shown between time and  ensemble averages
(except  at very  small values of $t/t_w$)  
 confirms the validity of the   time averaging procedure
 used to evaluate $r_{TRM}$.
 
Summarizing, the fast magnetization decay
occurring immediately after  field removal is  intermittent,
a further indication that all magnetization decay is intermittent 
and controlled by the magnitude of $r_{TRM}$. Secondly, time and ensemble averages
give similar estimates of the magnetization rate.

\section{Theory} 
As it transpires from our results, 
the main theoretical focus will  be on the rate of 
demagnetization $r_{TRM}$, which 
appears intimately related to the intermittent events. 
In   a record dynamics scenario~\cite{Sibani03,Sibani93a},  intermittent events, or \emph{quakes},
reflect significant configurational  changes which \emph{(i)} lead  from
one metastable configuration to another,   \emph{(ii)}  are irreversible 
and  \emph{(iii)} are triggered by extremal fluctuations. For thermal 
activated dynamics,  these would be thermal   fluctuations over record-sized energy  barriers. 
We presently assume that quakes  are statistically independent, 
and that measurable  effects    are all subordinated~\cite{Sibani06} to their occurrence.  
The time and age dependence of any  aging quantity 
is   then co-determined by  how many quakes  occur  between $t_w$ and $t$ 
 and by the   magnitude and nature of the  physical changes that  each quake entails.    
The number of quakes $n_I$ in the observation interval, $(t_w,t)$,  is a Poisson
distributed stochastic variable~\cite{Sibani03,Sibani93a} with  average 
 \begin{equation}
 \langle n_I(t_w,t)\rangle = \alpha \log(t/t_w). 
 \label{basic}
 \end{equation} 
The parameter $\alpha$ depends  linearly  
on the system size~\cite{Sibani06}, as it arises from 
an  extensive number of independent contributions from 
 different thermalized domains.  Furthermore,  
it is   temperature independent~\cite{Sibani05},  due    
to the implied self-similarity   of the energy landscape 
of  each thermalized domain~\cite{Sibani06b}. 
While the  above  properties are supposedly generic~\cite{Sibani03},   
 the  physical effects of the quakes  could be system and even variable  dependent.
However, even without a  specific  knowledge of  these effects,  
a subordination hypothesis  significantly  restricts
the  possible  age  dependencies~\cite{Sibani06}. 
of  physical quantities  of interest.
  
Subordination means that  any   time dependence is mediated 
 by  $n_I(t_w,t)$, whence 
 the TRM magnetization  becomes an (unknown)    stochastic process with  $n_I$ acting  
as an effective  `time'   variable.  In principle,    desired properties of  
the TRM can be found for fixed $n_I$,  and the underlying  time
dependencies can be re-introduced,  by averaging  $n_I$ according to the Poisson
distribution specified by Eq.~\ref{basic}. 

As quakes are seldom events,  we can treat their physical effects,
e.g. magnetization or energy changes,   as  statistically independent.
If the thermoremanent magnetization   is treated 
as a  Markov chain  with $n_I$ in the r\^{o}le of (discrete) time, 
its average $M_{TRM}(n_I)$   admits   an eigenvalue expansion~\cite{VanKampen92}. 
According to Eq.~\ref{basic}, the  range of $n_I$    will  
 be modest for  achievable time arguments and  
most terms in the aforementioned   eigenvalue expansion will    effectively 
remain constant.  To account for the   few  modes which  change during 
the  decay, we  tentatively write 
$ M_{TRM}(n_I) = c + c_0 \exp(a_0 n_I) + c_1 \exp(a_1 n_I) + c_2 \exp(a_2 n_I) + \ldots$
where $a_i$ are (real and negative) eigenvalues, all of order one or smaller.
Without loss of generality,  we can assume that for realistic values of $n_I$,
 $a_0 n_I$ remains  sufficiently  small to justify a 
 linear expansion of the first exponential, leading to
 \begin{equation}
 M_{TRM}(n_I) =
 c' + c_0  a_0 n_I  + c_1 \exp(a_1 n_I) + c_2 \exp(a_2 n_I) + \ldots 
 \end{equation}
Averaging  over $n_I$ yields 
\begin{equation}
M_{TRM}(t,t_w)   =   c' + c_0  a_0 \alpha \ln(\frac{t}{t_w} )  + c_1 (\frac{t}{t_w})^{\lambda_1} + c_2 (\frac{t}{t_w} )^{\lambda_2}+ \ldots
\label{mag_f}
\end{equation} 
where $\lambda_i = -\alpha ( 1-e^{a_i}) <0$ and $i =1,2$. 
Differentiating with respect to $t$, and re-arranging the terms,
 we finally obtain  
\begin{equation}
t \,\, r_{TRM}(t) - c_0  a_0 \alpha =   c_1 \lambda_1 
 ( \frac{t}{t_w})^{\lambda_1}+ c_2 \lambda_2 (\frac{t}{t_w} )^{\lambda_2}. 
\label{main_formula}
\end{equation}
The   above  expression has the same functional 
form as   Eq.~\ref{main_fit_formula}, with $a = c_0  a_0 \alpha$,
$b_1 = c_1 \lambda_1$ and  $b_2 = c_2 \lambda_2$.  
The weak $T$ dependence of both exponents and pre-factors should be expected as 
quakes are  exothermal\cite{Sibani06}.  The near $T$ independence 
of $a$ implies, banning unlikely cancellations, that $\alpha$ is itself temperature 
independent,  precisely as required by   record dynamics. 
 
All  physical observables   are    simply  related in   the  linear
 response regime~\cite{Nordblad86}.  
 E.g.\ the `relaxation rate', which in the present  notation is  
  given by   $S(t)= -\partial M_{TRM} /\partial \log (t-t_w) $,
can be cast into  the form $S(t_{obs})=- t_{obs}r_{TRM}(t_{obs}+t_w)$, where,  
 as we recall, $t_{obs} = t -t_w$. Inserting  Eq.~\ref{main_formula} for $r_{TRM}$ 
and plotting the outcome versus $\log_{10} t_{obs}$ reproduces  
the well known shape of $S(t)$:  a  broad maximum is present at  $t_{obs} = t_w$,
and a flat asymptotic value,  equal to $a$, is reached   for $t_{obs}\gg t_w$. 
The  asymptotically   logarithmic  TRM decay and, 
equivalently, the  constant asymptotic value of $S$ are widely observed
in complex systems,
To name a few cases,   they are seen in  switchable mirrors
after UV illumination~\cite{Lee05}, in  the  field-cooled magnetization 
of spin glasses~\cite{Mathieu01}, and in   the magnetic creep  of the ROM model~\cite{Nicodemi01}  
of magnetic flux creep in type II superconductors \cite{Oliveira05}. 
In the asymptotic regime, the rate is $r_{TRM} \propto 1/t$, and the time at 
$t_w$ at which the perturbation  is switched off is thus forgotten.  
Importantly,   the energy decay rate of the  EA spin
 glass~\cite{Sibani05} and of a p-spin model~\cite{Sibani06b}  
 is also (nearly) proportional the reciprocal of the age, $r_E(t) \propto 1/t$. 
(Recall  that, inconveniently, the symbol $t_w$ is used in 
Refs.~\cite{Sibani05,Sibani06}  in lieu of $t$.) 
In conclusion,  relaxation   in response to the initial thermal quench, which is never forgotten,
 seems to emerge  as  the main physical process
during  non-equilibrium aging, whether or not  an added field is present.
This is fully consistent  with   the subordination assumption used  
to interpret the present data.

\section{Summary and conclusions} 
Using   available spin-glass thermoremanent magnetization  data 
and a   general method of analysis, we have extracted 
the intermittent  properties of   the TRM decay
and interpreted them theoretically using record dynamics. 
In combination with a previous investigation of the configuration autocorrelation
function of the Edwards-Anderson spin-glass~\cite{Sibani06}, 
the present results suggests that conjugate autocorrelation and response
functions may inherit significant statistical properties from the quakes. This
can lead to  Fluctuation Dissipation-like relations~\cite{Cugliandolo97}  
between the two, even  in  an out-of-equilibrium situation, once the transient
effects of the field switch have died out.  
 
 The record dynamics  approach  melds   real  space features, 
i.e. the  independent  thermalized domains where
 quakes are initiated,  with configuration space features,
 i.e.the scale invariance of the energy landscape associated to  each domain,
  a concept repeatedly  stressed in hierarchical models~\cite{Dotsenko85,Sibani89,Joh96}.
Similar ideas have been applied to   `non-thermal' dynamics, e.g. 
memory effects  in   driven dissipative systems are linked to  the marginal stability 
of the attractors selected by  the dynamics~\cite{Coppersmith87,Tang87,Sibani93a}
and can also be understood by record dynamics.
By construction, the description becomes  invalid near
the    final equilibration time, since the postulate irreversibility
of quakes cannot be maintained.  
 
The complex interplay  between external noise 
and drift in  glassy  dynamics is a  pivotal   issue
in  non-equilibrium statistical physics,  and  many of its 
aspects are still only partially understood.
Further insights could  be obtained by 
applying the present  method to  linear response functions in
other situations, 
e.g. aging with  a  small  temperature step. The  intermittency of the 
heat loss has been studied in some detail in this situation~\cite{Sibani04a},
confirming that, in the Edwards-Anderson model,  the largest energy barrier overcome 
in the past evolution sets the time scale for future 
intermittent events.   
One  recent experimental  finding  is that  the TRM signal loses
its $t_w$ dependence in the extreme 
limit  $t_w \gg t$~\cite{Kenning06}.
As proposed in  that paper, the  `post-aging'
decay  is intrinsically related to the same 
mechanisms producing the usual aging effects, 
and is related to the memory of the initial state
set up by the cooling procedure. We  expect that  
experimental   investigations of intermittency in 
mesoscopic-scaled  systems will further 
elucidate this   and other memory effects in complex dynamics. 
 
\acknowledgments Support  from the Danish Natural Sciences Research Council
is gratefully acknowledged.   
\bibliography{../SD-meld}

\begin{thebibliography}{33}
\expandafter\ifx\csname natexlab\endcsname\relax\def\natexlab#1{#1}\fi
\expandafter\ifx\csname bibnamefont\endcsname\relax
  \def\bibnamefont#1{#1}\fi
\expandafter\ifx\csname bibfnamefont\endcsname\relax
  \def\bibfnamefont#1{#1}\fi
\expandafter\ifx\csname citenamefont\endcsname\relax
  \def\citenamefont#1{#1}\fi
\expandafter\ifx\csname url\endcsname\relax
  \def\url#1{\texttt{#1}}\fi
\expandafter\ifx\csname urlprefix\endcsname\relax\def\urlprefix{URL }\fi
\providecommand{\bibinfo}[2]{#2}
\providecommand{\eprint}[2][]{\url{#2}}

\bibitem[{\citenamefont{Alba et~al.}(1986)\citenamefont{Alba, Ocio, and
  Hammann}}]{Alba86}
\bibinfo{author}{\bibfnamefont{M.}~\bibnamefont{Alba}},
  \bibinfo{author}{\bibfnamefont{M.}~\bibnamefont{Ocio}}, \bibnamefont{and}
  \bibinfo{author}{\bibfnamefont{J.}~\bibnamefont{Hammann}},
  \bibinfo{journal}{Europhys. Lett.} \textbf{\bibinfo{volume}{2}},
  \bibinfo{pages}{45} (\bibinfo{year}{1986}).

\bibitem[{\citenamefont{{V.S. Zotev, G.F. Rodriguez, G.G. Kenning, R. Orbach,
  E. Vincent and J. Hammann}}(2003)}]{Zotev03}
\bibinfo{author}{\bibnamefont{{V.S. Zotev, G.F. Rodriguez, G.G. Kenning, R.
  Orbach, E. Vincent and J. Hammann}}}, \bibinfo{journal}{Phys. Rev. B}
  \textbf{\bibinfo{volume}{67}}, \bibinfo{pages}{184422}
  (\bibinfo{year}{2003}).

\bibitem[{\citenamefont{Rodriguez et~al.}(2003)\citenamefont{Rodriguez,
  Kenning, and Orbach}}]{Rodriguez03}
\bibinfo{author}{\bibfnamefont{G.~F.} \bibnamefont{Rodriguez}},
  \bibinfo{author}{\bibfnamefont{G.~G.} \bibnamefont{Kenning}},
  \bibnamefont{and} \bibinfo{author}{\bibfnamefont{R.}~\bibnamefont{Orbach}},
  \bibinfo{journal}{Phys. Rev. Lett.} \textbf{\bibinfo{volume}{91}},
  \bibinfo{pages}{037203} (\bibinfo{year}{2003}).

\bibitem[{\citenamefont{Jonason et~al.}(1998)\citenamefont{Jonason, Vincent,
  Hamman, Bouchaud, and Nordblad}}]{Jonason98}
\bibinfo{author}{\bibfnamefont{K.}~\bibnamefont{Jonason}},
  \bibinfo{author}{\bibfnamefont{E.}~\bibnamefont{Vincent}},
  \bibinfo{author}{\bibfnamefont{J.}~\bibnamefont{Hamman}},
  \bibinfo{author}{\bibfnamefont{J.~P.} \bibnamefont{Bouchaud}},
  \bibnamefont{and} \bibinfo{author}{\bibfnamefont{P.}~\bibnamefont{Nordblad}},
  \bibinfo{journal}{Phys. Rev. Lett.} \textbf{\bibinfo{volume}{81}},
  \bibinfo{pages}{3243} (\bibinfo{year}{1998}).

\bibitem[{\citenamefont{Sibani and Jensen}(2004)}]{Sibani04a}
\bibinfo{author}{\bibfnamefont{P.}~\bibnamefont{Sibani}} \bibnamefont{and}
  \bibinfo{author}{\bibfnamefont{H.~J.} \bibnamefont{Jensen}},
  \bibinfo{journal}{JSTAT} p. \bibinfo{pages}{P10013} (\bibinfo{year}{2004}).

\bibitem[{\citenamefont{Dotsenko}(1985)}]{Dotsenko85}
\bibinfo{author}{\bibfnamefont{V.~S.} \bibnamefont{Dotsenko}},
  \bibinfo{journal}{J. Phys. C} \textbf{\bibinfo{volume}{18}},
  \bibinfo{pages}{6023} (\bibinfo{year}{1985}).

\bibitem[{\citenamefont{Sibani and Hoffmann}(1989)}]{Sibani89}
\bibinfo{author}{\bibfnamefont{P.}~\bibnamefont{Sibani}} \bibnamefont{and}
  \bibinfo{author}{\bibfnamefont{K.~H.} \bibnamefont{Hoffmann}},
  \bibinfo{journal}{Phys. Rev. Lett.} \textbf{\bibinfo{volume}{63}},
  \bibinfo{pages}{2853} (\bibinfo{year}{1989}).

\bibitem[{\citenamefont{Joh and Orbach}(1996)}]{Joh96}
\bibinfo{author}{\bibfnamefont{Y.~G.} \bibnamefont{Joh}} \bibnamefont{and}
  \bibinfo{author}{\bibfnamefont{R.}~\bibnamefont{Orbach}},
  \bibinfo{journal}{Phys. Rev. Lett.} \textbf{\bibinfo{volume}{77}},
  \bibinfo{pages}{4648} (\bibinfo{year}{1996}).

\bibitem[{\citenamefont{K.H.~Hoffmann and Sibani}(1997)}]{Hoffmann97}
\bibinfo{author}{\bibfnamefont{S.~S.} \bibnamefont{K.H.~Hoffmann}}
  \bibnamefont{and} \bibinfo{author}{\bibfnamefont{P.}~\bibnamefont{Sibani}},
  \bibinfo{journal}{Europhys. Lett.} \textbf{\bibinfo{volume}{38}},
  \bibinfo{pages}{613} (\bibinfo{year}{1997}).

\bibitem[{\citenamefont{Jean-Philippe~Bouchaud and Vincent}(2001)}]{Bouchaud01}
\bibinfo{author}{\bibfnamefont{J.~H.} \bibnamefont{Jean-Philippe~Bouchaud},
  \bibfnamefont{Vincent~Dupuis}} \bibnamefont{and}
  \bibinfo{author}{\bibfnamefont{E.}~\bibnamefont{Vincent}},
  \bibinfo{journal}{Phys. Rev. B} \textbf{\bibinfo{volume}{65}},
  \bibinfo{pages}{024439} (\bibinfo{year}{2001}).

\bibitem[{\citenamefont{{H. Bissig, S. Romer, Luca Cipelletti Veronique Trappe
  and Peter Schurtenberger}}(2003)}]{Bissig03}
\bibinfo{author}{\bibnamefont{{H. Bissig, S. Romer, Luca Cipelletti Veronique
  Trappe and Peter Schurtenberger}}}, \bibinfo{journal}{PhysChemComm}
  \textbf{\bibinfo{volume}{6}}, \bibinfo{pages}{21} (\bibinfo{year}{2003}).

\bibitem[{\citenamefont{{L.~Buisson, L.~Bellon and
  S.~Ciliberto}}(2003)}]{Buisson03}
\bibinfo{author}{\bibnamefont{{L.~Buisson, L.~Bellon and S.~Ciliberto}}},
  \bibinfo{journal}{J. Phys. Cond. Mat.} \textbf{\bibinfo{volume}{15}},
  \bibinfo{pages}{S1163} (\bibinfo{year}{2003}).

\bibitem[{\citenamefont{{L. Buisson, M. Ciccotti, L. Bellon and S.
  Ciliberto}}(2004)}]{Buisson04}
\bibinfo{author}{\bibnamefont{{L. Buisson, M. Ciccotti, L. Bellon and S.
  Ciliberto}}}, in \emph{\bibinfo{booktitle}{Fluctuations and noise in
  materials}}, edited by \bibinfo{editor}{\bibfnamefont{M.~W.}
  \bibnamefont{D.~Popovic}} \bibnamefont{and}
  \bibinfo{editor}{\bibfnamefont{Z.}~\bibnamefont{Racz}}
  (\bibinfo{year}{2004}), pp. \bibinfo{pages}{150--163}.

\bibitem[{\citenamefont{Cipelletti and Ramos}(2005)}]{Cipelletti05}
\bibinfo{author}{\bibfnamefont{L.}~\bibnamefont{Cipelletti}} \bibnamefont{and}
  \bibinfo{author}{\bibfnamefont{L.}~\bibnamefont{Ramos}},
  \bibinfo{journal}{Journal of Physics: Condensed Matter}
  \textbf{\bibinfo{volume}{17}}, \bibinfo{pages}{R253} (\bibinfo{year}{2005}).

\bibitem[{\citenamefont{Sibani and Jensen}(2005)}]{Sibani05}
\bibinfo{author}{\bibfnamefont{P.}~\bibnamefont{Sibani}} \bibnamefont{and}
  \bibinfo{author}{\bibfnamefont{H.~J.} \bibnamefont{Jensen}},
  \bibinfo{journal}{Europhys. Lett.} \textbf{\bibinfo{volume}{69}},
  \bibinfo{pages}{563} (\bibinfo{year}{2005}).

\bibitem[{\citenamefont{{Paolo Sibani}}(2006)}]{Sibani06b}
\bibinfo{author}{\bibnamefont{{Paolo Sibani}}}, \bibinfo{journal}{Physical
  Review E} \textbf{\bibinfo{volume}{74}}, \bibinfo{pages}{031115}
  (\bibinfo{year}{2006}).

\bibitem[{\citenamefont{Sibani and Dall}(2003)}]{Sibani03}
\bibinfo{author}{\bibfnamefont{P.}~\bibnamefont{Sibani}} \bibnamefont{and}
  \bibinfo{author}{\bibfnamefont{J.}~\bibnamefont{Dall}},
  \bibinfo{journal}{Europhys. Lett.} \textbf{\bibinfo{volume}{64}},
  \bibinfo{pages}{8} (\bibinfo{year}{2003}).

\bibitem[{\citenamefont{Sibani and Littlewood}(1993)}]{Sibani93a}
\bibinfo{author}{\bibfnamefont{P.}~\bibnamefont{Sibani}} \bibnamefont{and}
  \bibinfo{author}{\bibfnamefont{P.~B.} \bibnamefont{Littlewood}},
  \bibinfo{journal}{Phys. Rev. Lett.} \textbf{\bibinfo{volume}{71}},
  \bibinfo{pages}{1482} (\bibinfo{year}{1993}).

\bibitem[{\citenamefont{Sibani}(2006)}]{Sibani06}
\bibinfo{author}{\bibfnamefont{P.}~\bibnamefont{Sibani}},
  \bibinfo{journal}{Europhys. Lett.} \textbf{\bibinfo{volume}{73}},
  \bibinfo{pages}{69} (\bibinfo{year}{2006}).

\bibitem[{\citenamefont{{Paul Anderson, Henrik Jeldtoft Jensen, L.P. Oliveira
  and Paolo Sibani}}(2004)}]{Anderson04}
\bibinfo{author}{\bibnamefont{{Paul Anderson, Henrik Jeldtoft Jensen, L.P.
  Oliveira and Paolo Sibani}}}, \bibinfo{journal}{Complexity}
  \textbf{\bibinfo{volume}{10}}, \bibinfo{pages}{49} (\bibinfo{year}{2004}).

\bibitem[{\citenamefont{{L.P. Oliveira, Henrik Jeldtoft Jensen, Mario Nicodemi
  and Paolo Sibani}}(2005)}]{Oliveira05}
\bibinfo{author}{\bibnamefont{{L.P. Oliveira, Henrik Jeldtoft Jensen, Mario
  Nicodemi and Paolo Sibani}}}, \bibinfo{journal}{Phys. Rev. B}
  \textbf{\bibinfo{volume}{71}}, \bibinfo{pages}{104526}
  (\bibinfo{year}{2005}).

\bibitem[{\citenamefont{Struik}(1978)}]{Struik78}
\bibinfo{author}{\bibfnamefont{L.}~\bibnamefont{Struik}},
  \emph{\bibinfo{title}{Physical aging in amorphous polymers and other
  materials}} (\bibinfo{publisher}{Elsevier Science Ltd}, \bibinfo{address}{New
  York}, \bibinfo{year}{1978}).

\bibitem[{\citenamefont{{Eric Vincent, Jacques Hammann, Miguel Ocio,
  Jean-Philippe Bouchaud, and Leticia F. Cugliandolo}}(1996)}]{Vincent96}
\bibinfo{author}{\bibnamefont{{Eric Vincent, Jacques Hammann, Miguel Ocio,
  Jean-Philippe Bouchaud, and Leticia F. Cugliandolo}}},
  \bibinfo{journal}{SPEC-SACLAY-96/048}  (\bibinfo{year}{1996}).

\bibitem[{\citenamefont{{Paolo Sibani, G.F. Rodriguez and G.G.
  Kenning}}(2006)}]{Sibani06a}
\bibinfo{author}{\bibnamefont{{Paolo Sibani, G.F. Rodriguez and G.G.
  Kenning}}}, \bibinfo{journal}{cond-mat/0601702}  (\bibinfo{year}{2006}).

\bibitem[{\citenamefont{Kampen}(1992)}]{VanKampen92}
\bibinfo{author}{\bibfnamefont{N.~G.~V.} \bibnamefont{Kampen}},
  \emph{\bibinfo{title}{Stochastic Processes in Physics and Chemistry}}
  (\bibinfo{publisher}{North Holland}, \bibinfo{year}{1992}).

\bibitem[{\citenamefont{Nordblad et~al.}(1986)\citenamefont{Nordblad,
  Svedlindh, Lundgren, and Sandlund}}]{Nordblad86}
\bibinfo{author}{\bibfnamefont{P.}~\bibnamefont{Nordblad}},
  \bibinfo{author}{\bibfnamefont{P.}~\bibnamefont{Svedlindh}},
  \bibinfo{author}{\bibfnamefont{L.}~\bibnamefont{Lundgren}}, \bibnamefont{and}
  \bibinfo{author}{\bibfnamefont{L.}~\bibnamefont{Sandlund}},
  \bibinfo{journal}{Phys. Rev. B} \textbf{\bibinfo{volume}{33}},
  \bibinfo{pages}{645} (\bibinfo{year}{1986}).

\bibitem[{\citenamefont{{M.~Lee P.~Oikonomou, P.~Segalova, T.~F.~Rosenbaum,
  A.~F.~Th.~Hoekstra and P.~B.~Littlewood}}(2005)}]{Lee05}
\bibinfo{author}{\bibnamefont{{M.~Lee P.~Oikonomou, P.~Segalova,
  T.~F.~Rosenbaum, A.~F.~Th.~Hoekstra and P.~B.~Littlewood}}},
  \bibinfo{journal}{J.Phys.:Condens. Matter} \textbf{\bibinfo{volume}{17}},
  \bibinfo{pages}{L439} (\bibinfo{year}{2005}).

\bibitem[{\citenamefont{{R.~Mathieu, P.~J\"{o}nsson, D.N.H.~Nam and P.
  Nordblad}}(2001)}]{Mathieu01}
\bibinfo{author}{\bibnamefont{{R.~Mathieu, P.~J\"{o}nsson, D.N.H.~Nam and P.
  Nordblad}}}, \bibinfo{journal}{Phys. Rev. B} \textbf{\bibinfo{volume}{63}},
  \bibinfo{pages}{092401} (\bibinfo{year}{2001}).

\bibitem[{\citenamefont{Nicodemi and Jensen}(2001)}]{Nicodemi01}
\bibinfo{author}{\bibfnamefont{M.}~\bibnamefont{Nicodemi}} \bibnamefont{and}
  \bibinfo{author}{\bibfnamefont{H.~J.} \bibnamefont{Jensen}},
  \bibinfo{journal}{J. Phys A} \textbf{\bibinfo{volume}{34}},
  \bibinfo{pages}{8425} (\bibinfo{year}{2001}).

\bibitem[{\citenamefont{{Leticia F. Cugliandolo, Jorge Kurchan and Luca
  Peliti}}(1997)}]{Cugliandolo97}
\bibinfo{author}{\bibnamefont{{Leticia F. Cugliandolo, Jorge Kurchan and Luca
  Peliti}}}, \bibinfo{journal}{Phys. Rev. E} \textbf{\bibinfo{volume}{55}},
  \bibinfo{pages}{3898} (\bibinfo{year}{1997}).

\bibitem[{\citenamefont{Coppersmith and Littlewood}(1987)}]{Coppersmith87}
\bibinfo{author}{\bibfnamefont{S.}~\bibnamefont{Coppersmith}} \bibnamefont{and}
  \bibinfo{author}{\bibfnamefont{P.}~\bibnamefont{Littlewood}},
  \bibinfo{journal}{Phys. Rev. B} \textbf{\bibinfo{volume}{36}},
  \bibinfo{pages}{311} (\bibinfo{year}{1987}).

\bibitem[{\citenamefont{Tang et~al.}(1987)\citenamefont{Tang, Wiesenfeld, Bak,
  Coppersmith, and Littlewood}}]{Tang87}
\bibinfo{author}{\bibfnamefont{C.}~\bibnamefont{Tang}},
  \bibinfo{author}{\bibfnamefont{K.}~\bibnamefont{Wiesenfeld}},
  \bibinfo{author}{\bibfnamefont{P.}~\bibnamefont{Bak}},
  \bibinfo{author}{\bibfnamefont{S.}~\bibnamefont{Coppersmith}},
  \bibnamefont{and}
  \bibinfo{author}{\bibfnamefont{P.}~\bibnamefont{Littlewood}},
  \bibinfo{journal}{Phys. Rev. Lett.} \textbf{\bibinfo{volume}{58}},
  \bibinfo{pages}{1161} (\bibinfo{year}{1987}).

\bibitem[{\citenamefont{{G.G. Kenning, G.F. Rodriguez and R.
  Orbach}}(2006)}]{Kenning06}
\bibinfo{author}{\bibnamefont{{G.G. Kenning, G.F. Rodriguez and R. Orbach}}},
  \bibinfo{journal}{Phys. Rev. Lett.} \textbf{\bibinfo{volume}{97}},
  \bibinfo{pages}{057201} (\bibinfo{year}{2006}).

\end{thebibliography}
\newpage
 
\subsubsection*{Caption for  Fig.1} 

(Color on line)  {\em (a)}: The circles show the  PDF of the magnetization change $\delta M$
for isothermal aging at  $T=0.83 T_g$, with the field cut  at $t_w=100$s.
 The values of $\delta M$ are  taken over short intervals    $\delta t=1.045$s
within  the observation  interval    $[1000, 4000]$. Diamonds show 
the PDF of $\delta M$  from the same data stream, 
but now taken over larger intervals    $\delta t=3 \times 1.045$s.
This leads to a much stronger intermittent component on the left wing of the PDF. 
The full lines  are  least square fits   of the  \emph{positive} values of $\delta M$ to a Gaussian  with zero average. 
  {\em (b)}: From the same data, the   average and  the variance 
  (the latter scaled as shown) of  $\delta M$'s  are plotted versus $\delta t$.
The dotted line is obtained by a  least square error  fit. 

\subsubsection*{Caption for Fig.2}

(Color on line) Panels 1-5: the main plots   show  the   estimated values 
of $ t \, \, r_{TRM}   -a $  versus $t/t_w$, for
$t_w= 50$ (right pointing triangles, unavailable for $T=0.6T_g$ and $T=0.9T_g$), 
$100$ (circles), $300$ (squares), $630$ (diamonds), $1000$ (pentagrams), 
$3600$ (hexagrams), $6310$ (asterisks)  and $(10000)$ (left pointing triangles).  
   The  lines are given by    Eq.~\ref{main_fit_formula}.
The inserts  compare  the    TRM decay measured at  $t_w=100$s (red line),  
with  the theoretical estimates (blue circles) obtained  by integrating the  
fitted   decay rate, see Eq.~\ref{finite_change}. The abscissa  $t/t_w-1$ is the ratio of the observation 
time to the field removal age. In   panel 6,  the parameter 
$a$ and the two exponents $\lambda_1$ and $\lambda_2$
are plotted vs. temperature. The lines are guides  to the eye.

\subsubsection*{Caption for Fig.3} 
(Color on line) The main  plot   shows  the unnormalized  PDF (circles) 
 of the quantity  $\delta M/\delta t -a$. The values of  $\delta M$  
are  collected    from  data streams taken with  different
 $t_w$ and  with  $T=0.83 T_g$. The denominator   $\delta t$ is the repeat time in each 
measurement, All data are taken within  intervals  $[t, 1.25t]$, where the value of  $t$ 
changes with  $t_w$, keeping a constant ratio $t/t_w=1.13$.   The  
parabola and the straight line 
are  fits to a Gaussian and an  exponential, respectively. The  insert 
 shows  the  $t/t_w$ dependence of  $t \, \, r_{TRM} t -a$ (dots), where  
 $r_{TRM}$  is  estimated  by 
 averaging  $\delta M/\delta t$ over the   ensemble of data points available at each
 $t/t_w$.   
  The full line, which   is  lifted from   the third
 plot of Fig.~\ref{big},  is  obtained using  a different  averaging procedure,
 and  is therefore   \emph{not} a direct fit to the data shown.  

\end{document}